\documentclass[12pt]{article}

\usepackage{amsmath}
\usepackage{latexsym}
\newcommand\be{\begin{equation}}
\newcommand\ee{\end{equation}}
\newcommand\bea{\begin{eqnarray}}
\newcommand\eea{\end{eqnarray}}
\newcommand\beas{\begin{eqnarray*}}
\newcommand\eeas{\end{eqnarray*}}

\usepackage[unicode=true,pdfusetitle,
 bookmarks=true,bookmarksnumbered=false,bookmarksopen=false,
 breaklinks=false,pdfborder={0 0 1},backref=false,colorlinks=false]
 {hyperref}

\def\tr{{\rm Tr}}

\def\Xint#1{\mathchoice
{\XXint\displaystyle\textstyle{#1}}%
{\XXint\textstyle\scriptstyle{#1}}%
{\XXint\scriptstyle\scriptscriptstyle{#1}}%
{\XXint\scriptscriptstyle\scriptscriptstyle{#1}}%
\!\int}
\def\XXint#1#2#3{{\setbox0=\hbox{$#1{#2#3}{\int}$ }
\vcenter{\hbox{$#2#3$ }}\kern-.5\wd0}}

\def\dashint{\Xint-}

\begin{document}

\title{De Alfaro, Fubini and Furlan from multi Matrix Systems}

%\author{\\
%Mthokozisi Masuku^1 and Jo\~ao P. Rodrigues^{1,2}      \footnote{Email: Joao.Rodrigues@wits.ac.za} 
%National Institute for Theoretical Physics \\
%School of Physics and Mandelstam Institute for Theoretical Physics \\
%University of the Witwatersrand, Johannesburg\\
%Wits 2050, South Africa \\
%Department of Physics, Brown University, Providence, RI 02912, USA}

\author{\\
Mthokozisi Masuku$^1$ and Jo\~ao P. Rodrigues$^{1,2}$\footnote{Email: joao.rodrigues@wits.ac.za} \\
\\
$^1$National Institute for Theoretical Physics \\
School of Physics and Mandelstam Institute for Theoretical Physics \\
University of the Witwatersrand, Johannesburg\\
Wits 2050, South Africa \\ \\
$^2$Department of Physics, Brown University \\ Providence, RI 02912, USA
}

%\begin{document}
\maketitle

\begin{abstract}
We consider the quantum mechanics of an even number of space indexed hermitian matrices. Upon complexification, we show that a closed subsector naturally parametrized by a matrix valued radial coordinate has a description in terms of non interacting $s$-state "radial fermions" with an emergent De Alfaro, Fubini and Furlan type potential, present only for two or more complex matrices. The concomitant $AdS_2$ symmetry is identified.The large $N$ description in terms of the density of radial eigenvalues is also described. 
\end{abstract}

\begin{flushright}
BROWN-HET-1685
\end{flushright}

\newpage

%\noindent
\section{Introduction}

The study of multi-matrix systems \footnote{By multimatrix systems we have in mind the integral or the quantum mechanics of a finite number of matrices}, and particularly their large $N$  \cite{'t Hooft:1973jz} limit is of great interest.  They are of interest in their own right as, for instance, in providing an equivalent reduced description of $QCD$  in terms of a finite number of matrices with quenched momenta \cite{Eguchi:1982nm}, \cite{Bhanot:1982sh}, \cite{Gross:1982at}, or, as is well known, and through the large $N$ limit of their description of D$0$ branes \cite{Polchinski:1995mt}, in providing a possible definition of $M$ theory \cite{Banks:1996vh}. 

%They arise in several compactifications of $\cal{N}=$$4$ SYM, such as on $S^{3}\times R$, from which a %plane-wave matrix theory \cite{Kim:2003rza} is obtained related to the $\cal{N}$$=4$ SYM dilatation operator  
%\cite{Beisert:2004ry}.

%Wigner
%In the context of the AdS/CFT duality \cite{Maldacena:1997re}, \cite{Gubser:1998bc}, \cite{Witten:1998qj},  
%due to supersymmetry and conformal invariance, correlators of supergravity and $1/2$ BPS states reduce to %calculation of free matrix model overlaps 
%\cite{Lee:1998bxa}, \cite{Corley:2001zk} or consideration of related matrix hamiltonians \cite{Berenstein:2004kk}. 
%For stringy states, in the context of the BMN limit \cite{Berenstein:2002jq} and $\cal{N}$ $=4$ SYM, similar %considerations apply 
%\cite{Constable:2002hw}, \cite{Beisert:2002ff}, \cite{deMelloKoch:2003pv}. 

In the context of the AdS/CFT correspondence \cite{Maldacena:1997re}, \cite{Gubser:1998bc}, \cite{Witten:1998qj},  they find many important applications.  
Due to supersymmetry and conformal invariance, correlators of supergravity and $1/2$ BPS states reduce to calculation of free matrix model overlaps 
\cite{Lee:1998bxa}, \cite{Corley:2001zk} or consideration of related matrix hamiltonians \cite{Berenstein:2004kk}. 
For stringy states, in the context of the BMN limit \cite{Berenstein:2002jq} and $\cal{N}$ $=4$ SYM, similar considerations apply \cite{Constable:2002hw}, \cite{Beisert:2002ff}, \cite{deMelloKoch:2003pv}. There is a precise phase space identification \cite{Donos:2005vm}  \cite{Rodrigues:2005ec} between the collective density description of the dynamics of a single matrix \cite{Jevicki:1979mb}, and the droplet description of the LLM \cite{Lin:2004nb} metric.  
Compactification of $\cal{N}$$=4$ SYM on $R \times S^3$ yields a plane-wave matrix theory \cite{Kim:2003rza}, related to the $\cal{N}$  $=4$ SYM dilatation operator  
\cite{Beisert:2004ry}. Multi-matrix, multi-trace operators with diagonal free two point functions have been identified 
\cite{Kimura:2007wy}, \cite{Bhattacharyya:2008rb}, allowing for the calculation of correlators beyond the planar limit. 

The AdS/CFT correspondence has in general provided many new insights into the properties of several different gravitational objects from their corresponding description in terms of  the large $N$ limit of of matrix valued (super) Yang-Mills theories. But it is also of great interest to understand how or if gravitational degrees of freedom emerge from the large $N$ limit of matrix theories, including in settings without supersymmetry or even conformal invariance.

It is this problem of "emergent geometry" that we investigate in this communication, in the context of the quantum mechanics of a even number $d=2m$ of space valued hermitian matrices. These can be thought of as D$0$ coordinates, although we will not consider a Yang-Mills potential, in effect restricting ourselves mostly to the free case. 

Geometries arise often with a high degree of symmetry, typically with spherical symmetry. It is an interesting question to ask if there is a sense in which a matrix valued "radial coordinate" can be defined. For an even number of hermitiean matrices (or an arbitrary number of complex matrices) a closed sub sector\footnote{In the sense of closure under Schwinger-Dyson equations} parametrized by a single matrix with the properties expected of such radial matrix has indeed been identified \cite{Masuku:2009qf} \cite{Masuku:2011pm}. 

The key result of this communication is to show that in this radial subsector a De Alfaro, Fubini and Furlan (dAFF) \cite{deAlfaro:1976je} $1/r^2$ potential emerges, with strength fixed and proportional to $N^2$. This is present only when the number of complex matrices $m$ is two or more. 

Single particle conformal quantum mechanics with a $1/x^2$ potential of arbitrary strength, the very context considered in the original dAFF work, has formed the basis of several studies of the $AdS_2/CFT_1$ correspondence \cite{AdS_2_1} \cite{Strominger:2003tm} \cite{AdS_2_2}, which has not yet been fully developed. 

We will establish in this article a "fermionic" description of the large $N$ ground state Hamiltonian in terms of $N$ non-interacting higher dimensional $s$-state radial fermions in the presence of a dAFF $1/r^2$ potential of well defined strength.  We identify the generators of the $SL(2,R)$ conformal group, isomorphic to the group of isometries of $AdS_2$.  

%This spirit if generalizes      
%We will follow the approach of \cite{Strominger:2003tm}, while generalizing it to this higher dimensional case, %and identify the generators of the $SL(2,R)$ conformal group, isomorphic to the isometries of $AdS_2$.       

It is hoped that the emergence of this conformal group in the radial sector of a higher dimensional system may shed further light into the  $AdS_2/CFT_1$ correspondence, which often arises in the product space of near horizon geometries of higher dimensional black-hole geometries, although this is not the purpose of this communication.

The paper is organized as follows: In Section 2, the matrix valued radial coordinate is identified, and the volume element for inner products of wavefunctions in the radial subsector is described. The fermionic description of this sector in terms of  the sum of $N$ noninteracting $d+1$ dimensional $s$-state single particle Hamiltonians is obtained in Section 3, and the emergence of a dAFF potential established. In Section 4, the generators of the $SL(2,R)$ conformal group are identified. Section 5 describes the large $N$ eigenvalue density description of the underlying Calogero model, which is then explicitly obtained for the $L_0$ $SO(2,1)$ generator. Section 6 is reserved for discussion and conclusions. The Appendix derives a technical result needed in Section 3.

\noindent
\section{Hamiltonian and Radial Sector}

We consider the Hamiltonian of $d=2m$ hermitean matrices $X_a$ lableled by a space index $a=1,...,2m$:

\begin{eqnarray}
		\hat{H} & = & -\frac{1}{2} \tr{ \sum_{a=1}^{2m}  \frac{\partial}{\partial M_a} \frac{\partial}{\partial M_a} } + V  \nonumber \\ & = & - \frac{ 1 }{ 2 } \nabla^{ 2 } + V. 
	 \label{alphab}
\end{eqnarray}
We complexify by introducing complex matrices:

$$
            Z_1= X_1+iX_2 \quad , Z_2= X_3+iX_4 \quad , \text{etc} .
$$

\noindent
Labeling these complex matrices  $Z_A ~,A=1,...,m $, we consider the matrix

\be\label{MatF}
\sum_{A=1}^m Z_A^{\dagger}Z_A ~.
\ee

\noindent
This matrix is hermitean and positive definite, and its eigenvalues

\be\label{eig}
\rho_i = r_i^2 ~,~ i=1,...,N ~,~ \rho_i \ge 0,
\ee

\noindent
have a natural interpretation as the eigenvalues of a matrix valued radial coordinate. 

One can consider a parametrization of the complex matrices $Z_A, A=1,..,m$ in terms of the matrix valued radial cordinate and $2m-1$ unitary matrices 
\footnote{For an explicit such parametrizations in the $m=1$ case, see \cite{Masuku:2009qf}.}. Although in this communication we will mostly discuss the laplacian operator, or the potential free case, considering potentials $V$ that depend only the radial matrix eigenvalues allow us to consistently restrict ourselves to radial wavefuntions, i.e., wavefunctions depending only on the eigenvalues (\ref{eig}). This sector has an enhanced 
$U(N)^{m+1}$ symmetry 

\be\label{sym}
              Z_A \to V_A Z_A V^{\dagger}  ~, ~ A=1,...,m.
\ee

%\noindent 
%and depends only on the radial eigenvalues. 

The measure in the innner product of two such wave function will then take the form

$$
\int \prod_A \prod_{ij} {{dZ_A}^\dagger}_{ij} {dZ_A}_{ij} = \int \prod_i {d\rho_i} {\cal{J}}(\rho_i) d[\text{Angular}],  
$$ 
with the "angular" degrees of freedom being integrated out.

${\cal{J}}(\rho_i)$ has recently been obtained in closed form \cite{Masuku:2011pm}. 
This results from the remarkable fact that correlators in this sector, with the enhanced symmetry (\ref{sym}), close under Schwinger-Dyson equations. 
The result is:

\bea
{\cal{J}}(\rho_i) &=&
%C_m  \prod_i d\rho_i \rho_i^{m-1} \prod_{i \ne j} \rho_i^{\frac{m-1}{2}} \rho_j^{\frac{m-1}{2}} |\rho_i - \rho_j| 
C_m  \prod_i d\rho_i\rho_i^{m-1} \prod_{i > j} \rho_i^{m-1} \rho_j^{m-1} (\rho_i - \rho_j)^2 \nonumber \\
&=&
D_m  \prod_i dr_i r_i^{2m-1} \prod_{i > j} r_i^{2m-2} r_j^{2m-2} (r_i^2 - r_j^2)^2 \\
&=&  C_m \prod_i d\rho_i\rho_i^{m-1} \Delta_{RM}^2(\rho_i) = D_m \prod_i dr_i r_i^{2m-1} \Delta_{RM}^2(r_i^2) \nonumber ,
\eea

\noindent
where the antisymmetric product
$$
\Delta_{RM}(\rho_i) \equiv \prod_{i > j}  \rho_i^{\frac{m-1}{2}} \rho_j^{\frac{m-1}{2}} (\rho_i - \rho_j) 
$$

\noindent
generalizes the well known Van der Monde determinant $\Delta = \prod_{i > j} (\rho_i - \rho_j)$, and $C_m$ and $D_m$ are numerical constants.

In the radial sector, the Laplacian then takes the form:
\begin{eqnarray}
- \frac{1}{2} \nabla^{ 2}_{ Radial } & = & - \frac{1}{2} \sum_{i} \frac{ 1 }{ \Delta_{RM}^2 ( r^{ 2}_{ i} )  } \frac{ 1 }{ r^{ 2m - 1 }_{ i } } \frac{ \partial }{ \partial r_{ i } } r^{ 2m - 1 }_{ i } \Delta_{RM}^2 ( r^{ 2}_{ i} ) \frac{ \partial }{ \partial r_{ i } }  \\
	& = & - 2 \sum_{i} \frac{1}{  \Delta_{RM}^2( \rho_{ i} ) } \frac{ 1 }{ \rho^{ m - 1 }_{ i } } \frac{ \partial }{ \partial \rho_{ i } } \rho^{ m }_{ i }\Delta_{RM}^2( \rho_{ i} ) \frac{ \partial }{ \partial \rho_{ i } } \nonumber,
\end{eqnarray}

The Hamiltonian acts on symmetric wavefunctions $\Phi( \rho_{i} )$ of the eigenvalues:
\begin{eqnarray}
\hat{H} \Phi( \rho_{i} ) & = & E \Phi( \rho_{i} ) \nonumber \\
\left( - \frac{1}{ 2 } \nabla^{ 2}_{ Radial } + {V}( \rho_{i} ) \right) \Phi( \rho_{i} ) & = & E \Phi( \rho_{i} ),
\end{eqnarray}

\section{Radial fermionic description and dAFF}

It is a well known result that in the singlet sector of a single hermitean matrix, the first quantized hamiltionian can be mapped to a system of non-interacting single particle fermions \cite{Brezin:1977sv}. This was shown to also be the case in the radial sector of the two (hermitean) matrix hamiltonian, which was mapped to a system of non interacting $2+1$ dimensional $s$-state single particle radial fermions \cite{Masuku:2009qf}. We show in this section that this result is true also in the general case of $m>1$ complex matrices but with the important difference that, in this case, an additional dAFF potential is induced.        

Define the anti-symmetric wavefunction $ \Psi( \rho_{i} ) $ as follows: 
$$
\Psi( \rho_{ i } ) \equiv \Delta_{RM}( \rho_{ i} ) \Phi( \rho_{i} ) 
$$

The Laplacian operator $\nabla^{ 2}_{ Radial }$ now acts on $ \Psi( \rho_{i} )$ as: 
$$
4 \sum_{i} \left( \frac{ 1 }{ \rho^{ m - 1 }_{ i } } \frac{ 1 }{ \Delta_{RM}( \rho_{ i} ) } \frac{ \partial }{ \partial \rho_{ i } } \Delta_{RM}( \rho_{ i} ) \right) \rho^{ m }_{ i } \left( \Delta_{RM}( \rho_{ i} ) \frac{ \partial }{ \partial \rho_{ i } } \frac{ 1 }{ \Delta_{RM}( \rho_{ i} ) } \right)\Psi( \rho_{i} ) 
$$

However, one has the identity:
\bea\label{fermident}
  &4& \sum_{i} \left( \frac{ 1 }{ \rho^{ m - 1 }_{ i } } \frac{ 1 }{ \Delta_{RM}( \rho_{ i} ) } \frac{ \partial }{ \partial \rho_{ i } } \Delta_{RM}( \rho_{ i} ) \right) \rho^{ m }_{ i } \left( \Delta_{RM}( \rho_{ i} ) \frac{ \partial }{ \partial \rho_{ i } } \frac{ 1 }{ \Delta_{RM}( \rho_{ i} ) } \right)\Psi( \rho_{i} ) \nonumber \\
&=&    \left( \sum_{i} \frac{4}{ \rho^{m-1}_{i} } \frac{ \partial }{ \partial \rho_{i} } \rho^{m}_{i} \frac{ \partial }{ \partial \rho_{i} } - \frac{{(N^2-1)(m-1)^2} }{ \rho_{i} }\right) \Psi( \rho_{ i} ) 
\eea
The proof of this identity is left to Appendix A. 

As a result, the Hamiltonian acting on $ \Psi( \rho_{i} ) $ now takes the form:

\bea
\left[ -   2\sum_{i} \frac{1}{ \rho^{m-1}_{i} } \frac{ \partial }{ \partial \rho_{i} } \rho^{m}_{i} \frac{ \partial }{ \partial \rho_{i} } + \frac{{(N^2-1)(m-1)^2} }{ 2\rho_{i} } + V ( \rho_{i} ) \right] \Psi( \rho_{ i} )  &=&  E \Psi( \rho_{i} ) , \nonumber \\
\left[ -   \frac{1}{2}\sum_{i} \frac{1}{ r^{2m-1}_{i} } \frac{ \partial }{ \partial r_{i} } r^{2m-1}_{i} \frac{ \partial }{ \partial r_{i} } + \frac{{(N^2-1)(m-1)^2}}{2 r_{i}^2 }  + V ( r_{i} ) \right] \Psi( r_{ i} )  &=&  E \Psi( r_{i} ) \nonumber 
\eea

This is now the sum of single particle $d+1=2m+1$ dimensional s-state hamiltonians, but with an additional radial dAFF potential. The coefficient of this $ \frac{1}{r^2}$ potential is uniquely determined. The fact that it is proportional to $N^2$ ensures that the system has a smooth 't Hooft limit. It is absent in the case of a single complex matrix. 

This first quantized hamiltonian acts on wavefunctions which are antisymmetric under the exchange of radial coordinates only, hence their referral to as radial fermions.   

\section{$AdS_2$}
It is well known that the conformal quantum mechanical hamiltonian  
$$
\hat{h} = \frac{1}{2} p^2 + \frac{\eta^2}{2x^2}
$$
has a $SL(2,R)$ conformal symmetry generated by $h$ together with \footnote{We use the conventions in \cite{Strominger:2003tm}, and consider our generators at $t=0$.}
$$
\hat{k}= \frac{x^2}{2} \, ; \qquad \hat{d} = \frac{1}{2}   (xp+px).
$$
The commutator algebra is:  
\be\label{calgebra}
   [\hat{d},\hat{h}]=2i\hat{h}  \qquad  [\hat{d},\hat{k}]=-2i\hat{k}  \qquad [\hat{h},\hat{k}]= -i\hat{d}
\ee
This is mapped to  $SO(2,1)$ generators in the standard way: 
$$
\hat{L}_0=\frac{1}{2} (\hat{h}+\hat{k}) \quad \hat{L}_{\pm1}=\frac{1}{2} (\hat{h}-\hat{k}\mp i\hat{d}),
$$
with algebra
$$
[\hat{L}_0,\hat{L}_{\pm 1}] = \pm \hat{L}_{\pm 1}; \quad [\hat{L}_{-1},\hat{L}_{1}]=2\hat{L}_0.
$$
 
In the fermionic picture, the single matrix hamiltonian with the same potential, and the other $SL(2,R)$ generators, can be written in terms of second quantized fields as:
\bea\label{SingleSecond}
\hat{H} &=& \int dx \Psi^{\dagger} (x) \left( \frac{1}{2} p^2 + \frac{\eta^2}{2x^2} \right) \Psi (x) \nonumber \\
\hat{K} &=& \int dx \Psi^{\dagger} (x)  \frac{x^2}{2} \Psi (x)  \\
\hat{D} &=&  \frac{1}{2}    \int dx \Psi^{\dagger} (x)   (xp+px) \Psi (x) \nonumber 
\eea
with
\be
\left\{ \Psi (x), \Psi^{\dagger} (x') \right\} = \delta(x-x')
\ee

We now generalize to the higher dimensional case. At the 1st quantized level ($d=2m$ $p_r=-i\partial_r$) one has:
\bea
\hat{h} (p_r, r)= \frac{1}{2} \frac{1}{r^{d-1}}p_r \, r^{d-1} \, p_r &+&  \frac{(N^2-1)(d-2)^2}{8r^2} \nonumber \\
= \frac{p_r^2}{2} - \frac{i (d-1)}{2r} p_r  &+&  \frac{(N^2-1)(d-2)^2}{8r^2} \nonumber \\
\hat{d} (p_r, r)=  r p_r -i \frac{d}{2} \, ;\quad 
& &\hat{k} (p_r, r)= \frac{r^2}{2} 
\eea
One can verify that the algebra (\ref{calgebra}) is satisfied. In terms of second quantized operators satisfying

\be\label{antiradial}
 \left\{  \Psi (r), \Psi^{\dagger} (r') \right\}= \frac{\delta(r-r')}{r^{d-1}},
\ee
one has as generators:

\bea
\hat{H} &=& \int r^{d-1} dr \Psi^{\dagger} (r) \hat{h} (p_r, r)  \Psi (r) \nonumber \\
\hat{K} &=& \int r^{d-1} dr \Psi^{\dagger} (r) \hat{k} (p_r, r)  \Psi (r) \\
\hat{D} &=& \int r^{d-1} dr \Psi^{\dagger} (r) \hat{h} (p_r, r)  \Psi (r) \nonumber
\eea

The simplest way to verify that these generators close the appropriate algebra, is to note that (\ref{antiradial}) suggests that we redefine:
$$
\tilde{\Psi} (r) \equiv r^{\frac{d-1}{2}} \Psi(r),  \quad \tilde{ \Psi}^{\dagger} (r') \equiv r^{\frac{d-1}{2}} \Psi^{\dagger} (r')
$$
This is also the redefinition of the fields in terms of which $p_r$ becomes explicitly hermitean. One finds:

\bea
\hat{K} &=& \int dr \tilde{ \Psi}^{\dagger} (r) \frac{r^2}{2}  \tilde{\Psi} (r) \nonumber \\
\hat{D} &=& \int dr \tilde{ \Psi}^{\dagger} (r) \frac{1}{2} (r p_r + p_r r) \tilde{\Psi} (r) \nonumber \\
\hat{H} &=& \int dr \tilde{ \Psi}^{\dagger} (r) (\frac{p_r^2}{2}  +  \frac{N^2(d-2)^2-1}{8r^2}) \tilde{\Psi} (r)
\eea

So the higher dimensional case has been mapped to a one-dimensional quantum mechanical conformal hamiltonian with 

$$
\eta^2 = \frac{1}{4} (N^2(d-2)^2-1)\, ,
$$
which has the required symmetry, as the quantum mechanical generators (\ref{SingleSecond}) close the conformal algebra for arbitrary $\eta$. 

This is the basis of the evidence of a possible emergence of an $AdS_2$ geometry, in a spirit similar to arguments presented in \cite{Strominger:2003tm}. 

\section{Density}

In this section, we describe the large $N$ description of the dynamics of eigenvalues in terms of their density. This provides a continuum description of the underlying Calogero model. We use well established collective field theory techniques \cite{Jevicki:1979mb}, which we apply to the radial sector of our Hamiltonian. Our results are new, and generalize previously obtained results for the single hermitean matrix, and also for two matrices.   For a single hermitean matrix, this approach corresponds to the bosonization of the fermionic description of the system \cite{Klebanov:1991qa}.

We recall the positive definite hermitian matrix defined from $m$ complex matrices $Z_A, A = 1, 2, 3,...,m$:
\begin{equation}
 \sum_{A} Z^{ \dag }_{A} Z_{A}. \nonumber
\end{equation}

Our collective field variables are defined as  
\begin{eqnarray}
\phi_{k} & = & \tr{ e^{ik \sum_{A} Z^{ \dag }_{A} Z_{A} } } = \sum_{i} e^{ ik r_{i}^{2} } = \sum_{i} e^{ ik \rho_{i} }; \nonumber \\
\phi(\rho) & = & \int{ d k e^{-ik\rho} \phi_{k} } = \sum_{i} \delta(\rho - r_{i}^{2} )  = \sum_{i} \delta(\rho - \rho_{i} ), \nonumber
\end{eqnarray}

The collective field construction is based on two operators (so called "joining" and "splitting" operators), which have been obtained previously \cite{Masuku:2009qf} \cite{Masuku:2011pm}:
\begin{eqnarray}
\Omega( \rho, \rho'; [ \phi ] ) & = & \int{ \frac{ dk' }{ 2 \pi } } \int{ \frac{dk }{ 2 \pi } e^{-ik \rho } e^{-ik' \rho' }  \sum_{A} \frac{ \partial \phi_{k} }{ \partial Z^{ \dag }_{A } } \frac{ \partial \phi_{k' } }{ \partial Z_{A } }  } \nonumber \\
 &=&  \partial_{ \rho} \partial_{ \rho' } \left[ \rho \phi( \rho ) \delta( \rho - \rho' ) \right],\nonumber \\
%\end{eqnarray}
%\begin{eqnarray}
	\omega( \rho; [ \phi ] ) & = & \int{ \frac{dk}{2 \pi } e^{-ik \rho } \frac{ \partial^{2} \phi_{k} }{ \partial Z^{ \dag }_{A} \partial Z_{A} } } \nonumber \\
	& = & \partial_{ \rho } \left(  \rho \phi( \rho )  \left[ 2 \dashint{ \frac{d \rho' \phi( \rho') }{ ( \rho - \rho' ) } + \frac{N(m - 1) }{ \rho } } \right] \right). \nonumber
\end{eqnarray}

The leading (in $N$) form of the collective field hamiltonian \cite{Jevicki:1979mb} then takes the form: 
\begin{eqnarray}
H_{coll}[\rho; [ \phi ] ] & = & 2 \int{ d \rho } \int{ d \rho' \Pi( \rho ) \Omega( \rho, \rho' ; [ \phi ] ) \Pi( \rho' ) }  \nonumber \\
	& + &  \frac{1}{2} \int{d \rho } \int{d \rho' \omega( \rho;[ \phi ] ) \Omega^{-1} ( \rho, \rho';[ \phi ] ) \omega( \rho'; [ \phi ] ) } + V[ \rho, [ \phi ] ] ,\nonumber \\
	& \equiv& \hat{H}_{K}[ \rho ;[\Pi, \phi ] ] + \Delta V[ \rho; [ \phi ] ] + V[ \rho; [ \phi ] ] ,\nonumber
\end{eqnarray}
where we have introduced the conjugate momentum $\Pi(\rho ) = \partial / i\partial \phi( \rho)$. Then

\begin{eqnarray}
	- \frac{1}{2} \nabla^{ 2}_{ Radial } & \to &\hat{H}_{K} + \Delta V \nonumber \\  &=& 2 \int{ d \rho \left( \partial_{ \rho }\Pi( \rho ) \right) \left[ \rho \phi( \rho) \right] \left( \partial_{\rho} \Pi( \rho ) \right) }  \nonumber \\
	& + &  \frac{1}{2} \int{ d \rho  \left( \rho \phi( \rho ) \right) \left[ 2 \dashint{ \frac{d \rho' \phi( \rho') }{ ( \rho - \rho' ) } + \frac{N(m - 1) }{ \rho } } \right]^{2} }\nonumber 
\end{eqnarray}
Some re-arrangment yields:
$$
\Delta V = 
2 \int_0^{\infty}  d \rho   \rho \phi( \rho ) 
\left[ \dashint_0^{\infty}{ \frac{d \rho' \phi( \rho') }{ ( \rho - \rho' ) } } \right]^2 + \frac{N^2(m-1)^2}{2} \int_0^{\infty}  d \rho \frac{ \phi( \rho )}{\rho}.
$$
In terms of the underlying Calogero eigenvalue system, this is equivalently written as: 
$$
\Delta V = 
2 \sum_i   \rho_i 
\left[ \sum_{j \neq i} \frac{1}{ ( \rho_i - \rho_j ) }  \right]^2 + \frac{N^2(m-1)^2}{2} \sum_i \frac{1}{\rho_i}
$$

We again find in addition to the usual repulsive term amongst the eigenvalues, the emergent dAFF $1 / \rho$ potential, which is present for $m>1$.  
  
When one extends to the whole line and define $\Phi(r) \equiv 2r\phi(r^{2}) = \Phi(-r)$, one can use the identity
\be
 \int^{ \infty}_{ -\infty }{ d r \Phi(r)
\left( \dashint^{ \infty}_{ -\infty }{ \frac{ d r' \Phi(r') }{ ( r - r' ) } } \right)^{2} }= \frac{ \pi^{2} }{3}  \int_{-\infty}^{\infty} dr \Phi^{3}(r) \nonumber
\ee
Then
\bea\label{localham}
- \frac{1}{2} \nabla^{ 2}_{ Radial } & \to &\hat{H}_{K} + \Delta V \nonumber \\
 &=& \frac{1}{2} \int_{0}^{\infty} dr \partial_r \Pi(r)\,\Phi(r)\, \partial_r \Pi(r) \\
    &+& \frac{ \pi^{2} }{ 6} \int^{ \infty }_{0 }{dr \Phi^{3}(r) } + \frac{N^{2} (m - 1)^{2}}{2} \int^{ \infty }_{0}{dr \left[ \frac{ \Phi(r) }{r^{2} } \right] } \nonumber
\eea
with
\be
[\Phi(r),\Pi(r')]=i\delta(r-r') \nonumber
\ee

It is worth while pointing out what (\ref{localham}) achieves: a continuous $r$ coordinate emerges, which combines naturally with the original time of quantum mechanics to yield a local theory. In this sense, a constructive holographic description has been achieved, which ultimately results from a change of variables in the gauge theory side, similarly to, and very much in the spirit of, the holographic construction of higher spin theories achieved in \cite{Koch:2010cy} from their $O(N)$ invariant duals.   

\subsection{$2\hat{L}_0$: Large $N$ background }
We now show how to obtain the large $N$ background associated with the $\hat{L}_0$ operator \footnote{For dimensional reasons we intriduce a mass parameter $w$}, 
$$
 2 \hat{L}_0= \hat{H}+\hat{K}  
$$
One has 
\begin{eqnarray}
 H_{coll} &=& \frac{1}{2} \int_{0}^{\infty} dr \partial_r \Pi(r)\,\Phi(r)\, \partial_r \Pi(r) \nonumber \\
 &+& \frac{ \pi^{2} }{ 6} \int^{ \infty }_{0}{dr \Phi^{3}(r) } + \frac{N^{2} (d-2)^{2}}{8} \int^{ \infty }_{0}{dr \left[ \frac{ \Phi(r) }{r^{2} } \right] } \nonumber \\
	& + & \frac{ \omega^{2} }{2} \int^{ \infty }_{0}{ d r \, r^2 \Phi( r ) } - \mu\left( N - \int^{ \infty}_{0}{dr \Phi(r) } \right) \nonumber
\end{eqnarray}
where the lagrange multiplier has been added to enforce the contraint $\int^{ \infty}_{0}{dr \Phi(r) }=N$.To make powers of $N$ explicit, we rescale
\begin{equation}
{r \rightarrow \sqrt{N} r } \hspace{8pt} {\Phi(r) \rightarrow \sqrt{N} \Phi(r)} \hspace{8pt} {\mu \rightarrow N \mu}; \hspace{8pt}\Pi(r) \to \Pi(r)/N \nonumber 
\end{equation}
We find
\be 
\quad V_{eff} \equiv \Delta V + V  \to N^2 V_{eff} \quad H_{K} \to H_{K}/N^2 \nonumber
\ee
The large $N$ is then the minimum of $V_{eff}$, and is easily shown to be: 
\begin{equation}
\Phi_0(r)  =  \frac{1}{\pi } \left( \frac{\omega}{2} \left( d-1 \right) - \omega^{2}r^{2} - \frac{ (d - 2)^{2} }{4} \frac{1}{r^{2} } \right)^{1/2} \hspace{20pt} r_{-} \leq r \leq r_{+} \nonumber
\end{equation}
\begin{equation}
r^{2}_{ \pm} = \frac{(d - 1)}{ 4 \omega} \pm \sqrt{ \frac{(d - 1)^{2} }{ 16\omega^{2} } - \frac{(d-2)^{2} }{ 4\omega^{2} } } \nonumber
\end{equation}

We observe that a Wigner distribution is present only for the single complex matrix ($m=1$), but that for two or more complex matrices the background has support within "hyper-annuli", or two "event horizons".

In \cite{Strominger:2003tm}, it has been argued that the diference between considering $\hat{H}$ versus $\hat{L}_0$ should be associated with different choices of time (Poincar\'e versus global time) on the gravity side. A discussion of the large $N$ $\hat{H}$ background in the context of such scenario, and possible identification of metric strucutures, is beyond the scope of this communication. 

It is interesting to observe that the presence of a dAFF potential plays an important role in the light front approach to the matter sector spectrum of QCD of \cite{Brodsky:2014yha}. Indeed the diference between the spectrum of $\hat{H}$ and that of $\hat{L}_0$ is related to a generated confining potential associated with $\hat{L}_0$.   

\section{Summary and Conclusions}
We considered in this communication the quantum mechanics of an even number of hermitean matrices in a restricted radial sector of the theory, and showed that a $1/r^2$ potential emerges with strength proportional to $N^2$, for sufficiently large number of matrices. This sector of the theory has a (radial) fermionic description in terms of higher dimentional single particle $s$-state hamiltonians in the presence of a dAFF potential. This generalizes previously known large $N$ fermionic descriptions of single hermitean matrices. $SL(2,R)$ conformal generators and generators of $SO(2,1)$ were identified, indicative if an $AdS_2$ geometry. The large $N$ eigenvalue density description of the underlying Calogero model was obtained, and explicitly obtained for the $L_0$ $SO(2,1)$ generator.      

Several issues of interest and for future investigation arise naturally from the results established in this communication. For instance, it is remarkable that the single particle hamiltonians and corresponding wave functions satisfy $AdS$ wave equations, despite the fact that one would not a priori associate the radial coordinate, as defined in this communication, with the holographic $z$ of an $AdS$ Poincar\'e patch. Also, there is a very natural metric associated with the large $N$ collective field theory description of the single hermitian matrix \cite{Demeterfi:1991tz}. It is of great interest to investigate if the analogue metric in the context of the results of this communication provides an explicit construction of an emergent $AdS$ metric. It is likely that this will require further understanding of the angular degrees of freedom (parametrizing $S_{d-1}$) \footnote{Generalizations of the dAFF system to include angular degrees of freedom have been considered in \cite{Bellucci:2002va}, for instance.}. A related issue is whether the original conformal quantum mechanical correlators can be obtained from generating functionals of gravity field sources \cite{Gubser:1998bc} \cite{Witten:1998qj}. These issues are currently being investigated. 

\section{Appendix}

We wish to prove the identity
\bea
  &4& \sum_{i} \left( \frac{ 1 }{ \rho^{ m - 1 }_{ i } } \frac{ 1 }{ \Delta_{RM}( \rho_{ i} ) } \frac{ \partial }{ \partial \rho_{ i } } \Delta_{RM}( \rho_{ i} ) \right) \rho^{ m }_{ i } \left( \Delta_{RM}( \rho_{ i} ) \frac{ \partial }{ \partial \rho_{ i } } \frac{ 1 }{ \Delta_{RM}( \rho_{ i} ) } \right)\nonumber \\
&=&    \left( 4\sum_{i} \frac{1}{ \rho^{m-1}_{i} } \frac{ \partial }{ \partial \rho_{i} } \rho^{m}_{i} \frac{ \partial }{ \partial \rho_{i} } - \frac{(N^2-1)(m - 1)^{2} }{ \rho_{i} } \right)  \nonumber
\eea
One has  
\be
		 \Delta_{RM}( \rho_{ i} ) \frac{ \partial }{ \partial \rho_{ i } } \frac{ 1 }{ \Delta_{RM}( \rho_{ i} ) } 
		 =  \left(  \frac{ \partial }{ \partial \rho_{ i } }- \frac{ a}{ \rho_{i} } -\sum_{k \neq i} \frac{ 1 }{ \rho_{i} - \rho_{k} } \right), \nonumber
\ee
where we have defined $a\equiv(N-1)(m-1)/2$. Similarly, 
\be
\left( \frac{1}{ \Delta_{RM}( \rho_{ i} ) } \frac{ \partial }{ \partial \rho_{ i } } { \Delta_{RM}( \rho_{ i} ) } \right)  =  
 \left( \frac{ \partial }{ \partial \rho_{ i } } + \frac{ a}{ \rho_{i} } +\sum_{k \neq i} \frac{ 1 }{ \rho_{i} - \rho_{k} }  \right). \nonumber
\ee
Defining $b = m-1$ for simplicity, one then has

\begin{eqnarray}
	& & \sum_{ i} \frac{ 4}{ \rho^{ b}_{i} } \left( \frac{ \partial }{ \partial \rho_{i} } + \frac{ a}{ \rho_{i} } +\sum_{k \neq \i} \frac{ 1 }{ \rho_{i} - \rho_{k} } \right)\rho^{ m }_{ i } \left( \frac{ \partial }{ \partial \rho_{i} } - \frac{ a}{ \rho_{i} } - \sum_{j \neq \i} \frac{ 1 }{ \rho_{i} - \rho_{j} }\right) = \nonumber \\
	& & \sum_{i} \frac{ 4 }{ \rho^{b}_{i} } \left( \frac{ \partial }{ \partial \rho_{ i } } \rho^{m}_{i} \frac{ \partial }{ \partial \rho_{ i } } - a\frac{ \partial }{ \partial \rho_{ i } } \rho^{b}_{i} - \frac{ \partial }{ \partial \rho_{ i } } \rho^{m}_{i} \sum_{ j \neq \i } \frac{ 1 }{ \rho_{i} - \rho_{j} } + a \rho^{b}_{i} \frac{ \partial }{ \partial \rho_{ i } } \right) + \nonumber\\
	& & \sum_{i} \frac{ 4 }{ \rho^{b}_{i} } \left( - a^{2} \rho^{b - 1}_{i} - a \sum_{ j \neq \i } \frac{ \rho^{b}_{i} }{ \rho_{i} - \rho_{j} } + \sum_{ k \neq \i } \frac{ \rho^{m}_{i} }{ \rho_{i} - \rho_{k} } \frac{ \partial }{ \partial \rho_{ i } } - \sum_{ k \neq \i } \frac{ { a } \rho^{b}_{i} }{ \rho_{i} - \rho_{k} }  \right) - \nonumber \\
	& & \sum_{i} \frac{ 4 }{ \rho^{b}_{i} } \left( \sum_{ k \neq \i } \sum_{ j \neq \i} \frac{ \rho^{m}_{i} }{ \rho_{i} - \rho_{k} } \frac{ 1 }{ \rho_{i} - \rho_{j} } \right). \nonumber
\end{eqnarray}

Except for the term $\frac{ \partial }{ \partial \rho_{ i } } \rho^{m}_{i} \frac{ \partial }{ \partial \rho_{ i } }$, one now acts with the derivatives. One finds that all terms linear in the derivative $\frac{ \partial }{ \partial \rho_{ i } }$ cancel out and one is left with: 
\begin{eqnarray}
	& = & 4 \sum_{i} \left( \frac{1}{ \rho^{b}_{i} } \frac{ \partial }{ \partial \rho_{i} } \rho^{m}_{i} \frac{ \partial }{ \partial \rho_{i} } - \frac{ab+a^2 }{ \rho_{i} }  - (m+2a) \sum_{ j \neq i } \frac{ 1 }{ \rho_{i} - \rho_{j} } \right)\nonumber \\
& + & 4 \sum_{i} \left( \sum_{ j \neq i } \frac{  \rho^{ m - b }_{ i } }{ (\rho_{i} - \rho_{j} )^{2} } 
	 - \sum_{k \neq i} \sum_{ j \neq i} \frac{ \rho^{m - b}_{i} }{ \rho_{i} - \rho_{k} } \frac{ 1 }{ \rho_{i} - \rho_{j} } \right) \nonumber \\
	& = & 4 \sum_{i} \left( \frac{1}{ \rho^{b}_{i} } \frac{ \partial }{ \partial \rho_{i} } \rho^{m}_{i} \frac{ \partial }{ \partial \rho_{i} } - \frac{ab+a^2 }{ \rho_{i} }  + \sum_{ j \neq i } \frac{  \rho_{ i } }{ (\rho_{i} - \rho_{j} )^{2} } - \sum_{k \neq i} \sum_{ j \neq i} \frac{ \rho_{i} }{ \rho_{i} - \rho_{k} } \frac{ 1 }{ \rho_{i} - \rho_{j} }\right). \nonumber 
\end{eqnarray}

The last two terms have already been encountered in \cite{Masuku:2009qf}, where it was already shown that they vanish. Indeed, one notices that: 
\begin{eqnarray}
\left( \sum_i  \sum_{k \neq i} \sum_{ j \neq i} \frac{ \rho_{i} }{ \rho_{i} - \rho_{k} } \frac{ 1 }{ \rho_{i} - \rho_{j} } -  \sum_i \sum_{ j \neq i } \frac{  \rho_{ i } }{ (\rho_{i} - \rho_{j} )^{2} }\right) & = & \sum_{i \neq j \neq k \neq i}  \frac{ \rho_{i} }{ \rho_{i} - \rho_{k} } \frac{ 1 }{ \rho_{i} - \rho_{j} }. \nonumber \\
\end{eqnarray}
This is easily seen to vanish by considering any three distinct eigenvalues. 

Hence, we have established the result (\ref{fermident}):  
\begin{eqnarray}
& & 4 \sum_{i} \left( \frac{ 1 }{ \Delta_{RM}( \rho_{ i} ) } \frac{ 1 }{ \rho^{ m - 1 }_{ i } } \frac{ \partial }{ \partial \rho_{ i } } \Delta_{RM}( \rho_{ i} ) \right) \rho^{ m }_{ i } \left( \Delta_{RM}( \rho_{ i} ) \frac{ \partial }{ \partial \rho_{ i } } \frac{ 1 }{ \Delta_{RM}( \rho_{ i} ) } \right) \nonumber \\
	&= & 4 \sum_{i} \left( \frac{1}{ \rho^{b}_{i} } \frac{ \partial }{ \partial \rho_{i} } \rho^{m}_{i} \frac{ \partial }{ \partial \rho_{i} } - \frac{ab+a^{2} }{ \rho_{i} } \right) \nonumber \\
	& = & \sum_{i} \left( \frac{4}{ \rho^{m-1}_{i} } \frac{ \partial }{ \partial \rho_{i} } \rho^{m}_{i} \frac{ \partial }{ \partial \rho_{i} } - \frac{{(N^2-1)(m-1)^2} }{ \rho_{i} }\right). \nonumber
\end{eqnarray}

\section{Acknowledgements}
We would like to thank Robert de Mello Koch and Antal Jevicki for interest in this work and for comments.
J.P.R. would like to thank the hospitality extended to him by the Theory High Energy Group of Brown University during a visit, when most of this article was written up. This research is supported in part by the National Research Foundation of South Africa (unique grant number 93440).

\end{document}